%% file: nips_2016_v3.tex
\documentclass{article}

\PassOptionsToPackage{square, numbers}{natbib}

 \usepackage[final]{nips_2016}

 \usepackage{threeparttable}
\usepackage{paralist}

\usepackage[utf8]{inputenc} 
\usepackage[T1]{fontenc}    
\usepackage{hyperref}       
\usepackage{url}            
\usepackage{booktabs}       
\usepackage{amsfonts}       
\usepackage{nicefrac}       
\usepackage{microtype}      

\usepackage{color}
\usepackage{color,graphicx,amssymb,amsmath,amsthm}

\usepackage{lipsum}

\newcommand{\tom}[1]{}
\newcommand{\itk}{^p}
\newcommand{\kp}{^{p+1}}
\newcommand{\eqhd}{\text{hard}}

\input{definitions}

\newcommand*{\affaddr}[1]{#1} 
\newcommand*{\affmark}[1][*]{\textsuperscript{#1}}

\title{ 
Non-negative Factorization of the \\ Occurrence Tensor from Financial Contracts
}

\author{%
Zheng Xu\affmark[1],\  Furong Huang\affmark[3], \ Louiqa Raschid\affmark[2],\  Tom Goldstein\affmark[1]\\
\affaddr{\affmark[1]Department of Computer Science, \affmark[2]Smith School of Business, University of Maryland
}\\
\affaddr{\affmark[3]Microsoft Research, New York City}\\
}

\begin{document}

\maketitle

\vspace{-0.5cm}
\begin{abstract}
\small

We propose an algorithm for the non-negative factorization of an occurrence tensor built from heterogeneous networks.  
We use $\ell_0$ norm to model sparse errors over discrete values
(occurrences), and use decomposed factors to model the embedded groups of nodes. 
\tom{This is vague.  Add another sentence or two making it clear what problems, and what applications, you're attacking}  
An efficient splitting method is developed to optimize the nonconvex and nonsmooth objective. We study both synthetic problems and a new dataset built from financial documents, resMBS.
\end{abstract}


\vspace{-0.5cm}
\section{Introduction}
\label{sec:intro}
Tensor factorization is a powerful approach to a myriad of unsupervised learning problems
\cite{anandkumar2014tensor,kolda2009tensor,papalexakis2016tensors}.
\tom{What does NOTF stand for?  What it is?  A tensor?  A factorization?  I can't tell} 
We propose an efficient algorithm called non-negative occurrence tensor factorization (NOTF) for analyzing heterogeneous networks.
As an application of NOTF, we study the resMBS dataset\cite{burdick2016dsmm,xu2016dsmm}, which consists of the relationships that financial institutions (FIs, e.g., Bank of America) play roles (e.g., issuer) in financial contracts (FCs). ResMBS is automatically extracted from a collection of public financial contracts 
.  We can represent resMBS as a three-mode  (FC, FI, Role) 
\tom{*briefly* describe what these features represent } 
occurrence tensor with non-negative discrete values corresponding to the confidence of the relationship extraction.
\tom{confidence in what?}

The occurrence tensor has several characteristics. 
First, the tensor has positive discrete values. 
Second, the observed tensor contains sparse noise 
corresponding to errors 
caused by inaccurate extraction. 
\tom{Errors due to what?}
Most importantly, we expect that FIs will play specific roles across
multiple FCs to create a community.
The observed occurrence tensor could then be decomposed 
into a low-rank tensor together with sparse noise.
This decomposition can be used to clean and complete the observation, and
also allow us to understand the underlying (FC, FI, Role) communities.


The proposed NOTF\tom{Write out what NOTF stands for, the  put (NOTF) in parentheses.  You need to define the term the first time you use it - even if you defined it in the abstract.}\zheng{appear in first paragraph}
extends the 
CANDECOMP/PARAFAC (CP) 
\tom{What does this stand for?} 
decomposition \cite{carroll1970analysis,harshman1970foundations} 
of a tensor. After decomposing the occurrence tensor, each 
rank-one tensor component represents a (FC, FI, Role) community.
Hence the decomposed factors are constrained to be non-negative. Instead of the $\ell_2$ norm for factorization of real valued tensors with Gaussian noise,  the $\ell_0$ norm is considered for the discrete values and the sparse errors. We develop 
 an algorithm based on the alternating direction method of multipliers (ADMM) \cite{boyd2011admm,xu2016adaptive} 
 to optimize the nonconvex and nonsmooth objective of NOTF. 

We briefly discuss some related research.
While tensor decomposition dates back to the 1920s,
robust tensor decomposition solutions have been presented in some 
recent papers 
\cite{kolda2009tensor,anandkumar2015tensor,goldfarb2014robust,gu2014robust}. 
Standard CP decomposition has been applied to heterogenous networks
\cite{maruhashi2011multiaspectforensics}. 
Discrete valued tensors have been studied in \cite{schein2015bayesian}. 
ADMM has been extensively used as a solver 
for robust tensor recovery \cite{goldfarb2014robust,gu2014robust}, 
and also for standard non-negative tensor factorization\cite{liavas2015parallel}. 
Non-$\ell_2$ norms have been used for CP decomposition of
real valued tensors in computer vision \cite{huang2008robust,chen2016robust}. 
ADMM for $\ell_0$ norm was empirically studied in \cite{xu2016empirical}.
Probabilistic communities were discussed for financial documents in \cite{xu2016dsmm}.

\section{Non-negative Occurrence Tensor Factorization}
\label{sec:ngtf}
We use notations similar to \cite{kolda2009tensor}. Vectors and matrices are denoted by lowercase and capital letters, respectively. Higher mode tensors are denoted by Euler script letters. Three mode tensors $\tsX \in \bbR^{N_1 \times N_2 \times N_3} $ are used as an example in this paper. Fibers are column vectors extracted from tensors by fixing every index but one, e.g., $x_{:jk}, x_{i:k}, x_{ij:}$. Mode-d fibers are arranged to get the mode-d unfolding matrix $X_{(d)}$, e.g., $X_{(1)} = [x_{:jk}] \in \bbR ^{N_1 \times N_2N_3}$. Denote the vector outer product by $\circ$, then the CP decomposition factorizes a tensor into a sum of rank-one tensors as $\tsX = \sum_{r=1}^R a_r \circ b_r \circ c_r$. If we denote $A=[a_r] \in \bbR^{N_1 \times R}$, $B=[b_r] \in \bbR^{N_2 \times R}$ and $C=[c_r]\in \bbR^{N_3 \times R}$,  the Kronecker product $A \otimes B \in \bbR^{N_1N_2 \times RR}$, and the Khatri-Rao product $A \odot B = [a_r \otimes b_r] \in \bbR^{N_1N_2 \times R}$, then we can compactly represent CP as $\tsX = \sum_{r=1}^R a_r \circ b_r \circ c_r = \tscp{A,B,C}$, which is equivalent to the unfolding representation   $X_{(1)} = A(C \odot B)\trans$, $X_{(2)} = B(C \odot A)\trans$, and $X_{(3)} = C(B \odot A)\trans$. 

We are seeking to recover a tensor $\tsX = \tscp{A,B,C}$ that is at most rank $R$ from the noisy observations $\tsO$.  Each rank-one tensor $a_r \circ b_r \circ c_r$ captures a community and $a_r \geq 0, b_r \geq 0, c_r \geq 0$ represent the weights of nodes (e.g., FI, FC, and roles) in the community. For discrete valued tensors, we minimize the sparse error measured with the $\ell_0$ norm rather than the $\ell_2$ loss in standard decomposition, 
\begin{align}
\min_{A,B,C} \| \tscp{A,B,C} - \tsO \|_0, \ \st \ A \geq 0, B \geq 0, C \geq 0, \label{eq:prob}
\end{align} 
where $\| \tsX \|_0 = \sum_{i,j,k} 1\!\!1_{\{z:\, | z| > 0\}}(x_{ijk})$ is the counts of nonzero values in a tensor, $1\!\!1_{S}$ is the indicator function of the set $S$: $1\!\!1_{S}(v) = 1$, if $v\in S$, and $1\!\!1_{S}(v) = 0$, otherwise. 

We minimize the NOTF objective \eqref{eq:prob} by introducing an intermediate variable $\tsU$, 
\begin{align}
\min_{\tsU, A,B,C} \| \tsU  \|_0 + \iota_{ \{z:\, z \geq 0 \} } (A,B,C) , \ \st  \ \tsU = \tscp{A,B,C}-\tsO, \label{eq:prob2}
\end{align} 
where $\iota_{S}$ is the characteristic function of the set $S;$ $\iota_{S}(v) = 0$, if $v\in S$, and $\iota_{S}(v) = \infty$, otherwise. We then apply alternating direction method of multipliers (ADMM) \cite{goldstein2009split,boyd2011admm, goldstein2014fast, xu2016adaptive} 
by introducing dual variables $\gl$ and alternatively solving subproblems of $\tsU$ and $A,B,C$, with $p$ indexing iterations,
\begin{align}
\tsU\kp & = \arg\min_{\tsU} \| \tsU \|_0 + \frac{\tau}{2} \big\| \tsU - \tscp{A\itk, B\itk, C\itk} +\tsO + \gl\itk \big\|_F^2  \label{eq:admm1}\\
A\kp, B\kp, C\kp & = \arg\min_{A, B, C} \iota_{ \{ z:\, z \geq 0 \} } (A,B,C)  + \frac{\tau}{2} \big\| \tsU\kp - \tscp{A, B, C} +\tsO + \gl\itk \big\|_F^2 \label{eq:admm2}\\
\gl\kp &= \gl\itk +  \tsU\kp - \tscp{A\kp, B\kp, C\kp} +\tsO, \label{eq:admm3}
\end{align}
where $\| \tsX \|_F = \sqrt{\sum_{i,j,k}  x_{ijk}^2}$, and $\tau$ is a hyperparameter called the penalty parameter.

Subproblem \eqref{eq:admm1} can be solved by the proximal operator of the $\ell_0$ norm, known as hard-thresholding,
\begin{align}
\tsU\kp & = \eqhd(\tscp{A\itk, B\itk, C\itk} -\tsO- \gl\itk, \, 1/\tau),
\end{align}
where $\eqhd(\tsZ, t) = \arg\min_\tsX \|\tsX\|_0 + \nicefrac{1}{2t} \| \tsX - \tsZ \|_F^2  =  \tsZ * \mcI_{\{z:|z| > \sqrt{2t}\}}(\tsZ)$, with $*$ representing the element-wise Hadamard product,  and $\mcI_{S}(\tsX) = [1\!\!1_{S}(x_{ijk})]$ r the element-wise indicator function.

Subproblem \eqref{eq:admm2} is nonnegative tensor factorization, which can be solved by alternatively optimizing one of $A,B,$ or $C$ when the other two are fixed, with $q$ indexing iterations,
{\small
\begin{align}
& A^{p, q+1}  =  \arg\min_{A} \iota_{ \{ z:\, z \geq 0 \} } (A)  + \frac{\tau}{2} \| (\tsU\kp+\tsO + \gl\itk)_{(1)} -   A(C^{p, q} \odot B^{p, q})\trans \|_F^2 \label{eq:sub2start} \\
 & \,\ =   \max\{ (\tsU\kp+\tsO + \gl\itk)_{(1)} (C^{p, q} \odot B^{p, q}) ((C^{p,q})\trans(C^{p,q}) * (B^{p,q})\trans(B^{p,q}))^{\dagger}, \, 0\}\\
& B^{p, q+1}  =  \arg\min_{B} \iota_{ \{ z:\, z \geq 0 \} } (B)  + \frac{\tau}{2} \| (\tsU\kp+\tsO + \gl\itk)_{(2)} -   B(C^{p, q} \odot A^{p, q+1})\trans \|_F^2 \\
& \,\ =  \max\{ (\tsU\kp+\tsO + \gl\itk)_{(2)} (C^{p, q} \odot A^{p, q+1}) ((C^{p, q})\trans(C^{p, q}) * (A^{p,q+1})\trans(A^{p,q+1}))^{\dagger}, \, 0\}\\
& C^{p, q+1}  =  \arg\min_{C} \iota_{ \{ z:\, z \geq 0 \} } (C)  + \frac{\tau}{2} \| (\tsU\kp+\tsO + \gl\itk)_{(3)} -   C(B^{p,q+1} \odot A^{p,q+1})\trans \|_F^2 \\
 & \,\ =   \max  \{ (\tsU\kp+\tsO + \gl\itk)_{(3)} (B^{p,q+1} \odot A^{p,q+1}) ((B^{p,q+1})\trans(B^{p,q+1}) * (A^{p,q+1})\trans(A^{p,q+1}))^{\dagger}, \, 0\}. \label{eq:sub2end}
\end{align}
}

\noindent Each subproblem is a constrained least squares problem the recovers the mode-d unfolding matrix  $(\tsU\kp+\tsO + \gl\itk)$, starting from $A^{p, 0} = A^{p}, B^{p, 0} = B^{p}, C^{p, 0} = C^{p}$,  and updating $A,B,C$ according to \eqref{eq:sub2start}-\eqref{eq:sub2end} until convergence, then $A^{p+1} = A^{p, \scriptsize{\text{end}} }, B^{p+1} = B^{p, \scriptsize{\text{end}}}, C^{p+1} = C^{p,\scriptsize{\text{end}}}$. The updates \eqref{eq:sub2start}-\eqref{eq:sub2end} for the subproblems \eqref{eq:admm2} usually converge in less than ten iterations when warm started from the previous iteration. Relative ``residuals'' are used to monitor the convergence of  \eqref{eq:admm1}-\eqref{eq:admm3}, and are defined by
\begin{align}
\text{res}_1 &= \frac{\big\| \tscp{A^{p, q}, B^{p,q}, C^{p,q}} -\tscp{A^{p, q-1}, B^{p,q-1}, C^{p,q-1}} \big\|_F } {\big\| \tscp{A^{p, q-1}, B^{p,q-1}, C^{p,q-1}} \big\|_F } 
\end{align}
\begin{align}
\text{res}_2 &= \max \left\{ \frac{\big\| \tscp{A^{p}, B^{p}, C^{p}} -\tscp{A^{p-1}, B^{p-1}, C^{p-1}} \big\|_F } {\big\| \tscp{A^{p-1}, B^{p-1}, C^{p-1}} \big\|_F }, \,  \frac{\| \gl\itk - \gl^{p-1} \|_F } {\| \gl^{p-1} \|_F }\right\}.
\end{align}
The algorithm converges when $\text{res}_1 < \gge$ and $\text{res}_2 < \gge$, with typical $\gge = 10^{-3}$. The relative residual $\text{res}_2$ is inspired by the primal and dual residuals in \cite{boyd2011admm}. Note that we are not seeking a {\em unique} decomposition of $A,B,C$, but rather a stable low rank construction $\tscp{A,B,C}$ of observation $\tsO$ that has minimum sparse error. 

\section{Experiments on Synthetic Data}

We test the NOTF algorithm on a synthetic dataset constructed as follows: 
\begin{inparaenum}[(1)]
\item We create random sparse matrices $A \in \bbR^{50 \times 3}, B \in \bbR^{20 \times 3}, C \in \bbR^{10 \times 3}$, 
where the sparse ratios (ratio of zero to non-zero values) are $70.67\%, 55\%$ and $30\%$, respectively. 
Each nonzero value is uniformly sampled from range $(0, 1)$. 
\item Create the ground truth low rank matrix $\tsX = \mcI_{\{z: |z|> 0\}}(\tscp{A,B,C}) \in \bbR^{50 \times 20 \times 10}$;
it is the indicator tensor of nonzero values in the CP reconstruction from $A,B,C$; $\tscp{A,B,C}$ has CP rank $3$.
Note that $\tsX$ is a discrete (binary) valued tensor with sparsity ratio  $75.75\%$. 
\item Create the observation tensor $\tsO$ by adding noise that flips a small portion of the binary values in $\tsX$.   
\end{inparaenum}

NOTF reconstructs a low rank matrix $\tscp{\hat A,\hat B, \hat C}$ with at most CP rank $R$, 
where $A \in \bbR^{50 \times R}, B \in \bbR^{20 \times R}, C \in \bbR^{10 \times R}$. 
We vary the ratio of noise, i.e., the percentage of flips in $\tsO$, up to 10\%, and the rank 
parameter $R$ up to 10.
We report on convergence iterations, false positive and false negative counts when reconstructing 
the binary values in tensor $\tsX$ and $\tsO$, and the mean square error for reconstructing $\tsX$ and $\tsO$.
We compare NOTF with $\ell_0$ norm in \eqref{eq:prob} with non-negative tensor factorization (NTF) 
baselines using the $\ell_1$ norm and $\ell_2$ norms. 
Both NOTF and baseline methods 
are initialized with a CP decomposition of the 
observation tensor $\tsO$, $\gl^0 = 0$, with penalty parameter $\tau = 10$,
and implementated in Matlab using the Tensor toolbox \cite{TTB_Dense}.

\begin{figure}[t]
\centerline{
\includegraphics[width=\linewidth]{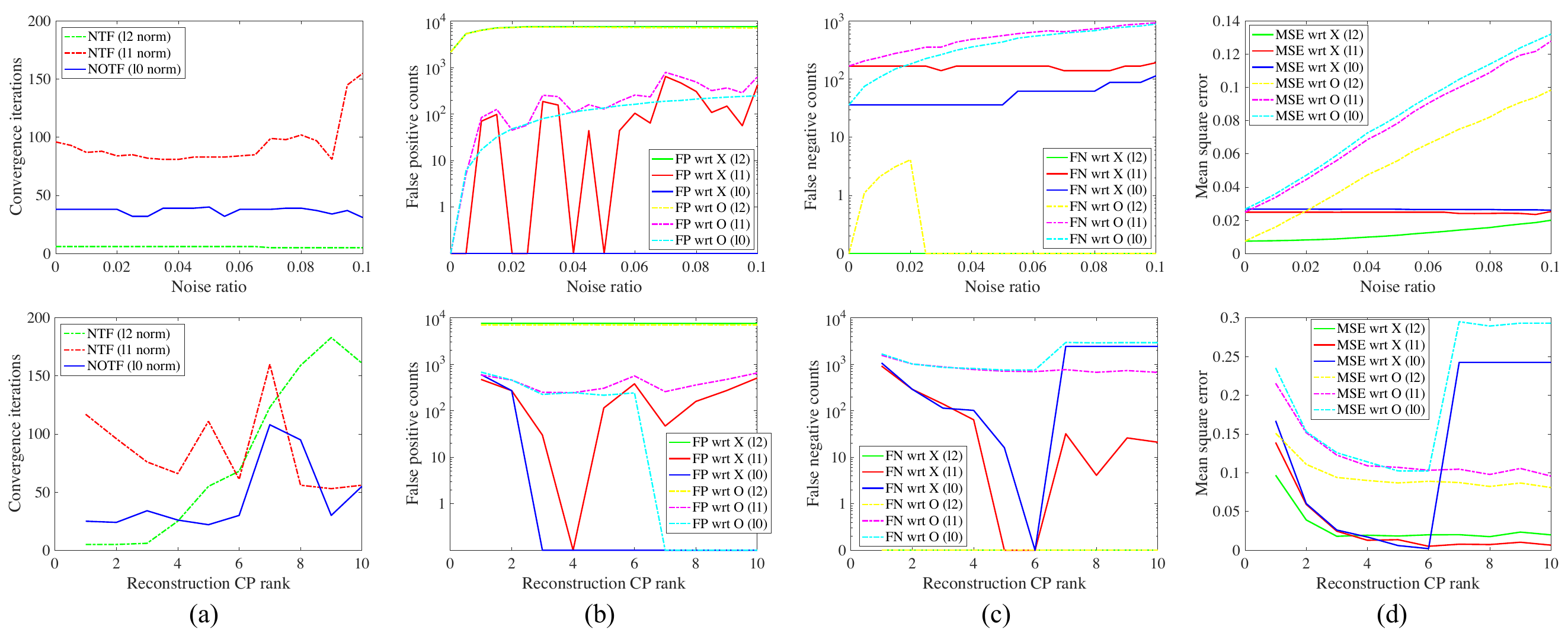}
}
\caption{\small  (a) Convergence iteration, (b) false positive count, (c) false negative count, and 
(d) mean square error when varying the noise ratio (top) and varying the tensor CP rank $R$ for NOTF (bottom)
for the synthetic dataset.  We set rank $R=3$ when varying the noise ratio (top) and set a noise ratio 
of $10\%$ when varying the CP rank (bottom). 
Note that reconstruction errors with respect to both groundtruth $\tsX$ and noisy observation $\tsO$ 
are presented in (b)-(d). 
}
\label{fig:noise}
\end{figure}

\cref{fig:noise} presents the results of varying the noise ratio up to 10\% (top) with rank parameter $R=3$, 
and varying the rank parameter $R$ up to 10 (bottom) with noise of 10\%, for the synthetic dataset.
We observe that the proposed NOTF solution with $\ell_0$ norm 
performs well
on the 
discrete measures (false positive  and false negative counts in (b) and (c));
we consider non-zeros as positives and zeros as negatives in a tensor.
NOTF achieves zero false positives and a relatively low false negative count over all noise ratios.
The $\ell_2$ baseline achieves zero false negatives, but the false positive counts are quite large.
The $\ell_1$ baseline achieves larger errors than NOTF on both positives and negatives, and is 
slower to converge. 

\cref{fig:noise} (d) shows that NOTF with $\ell_0$ norm does not outperform the baselines 
for mean square error; this is not surprising. 
The discrete measurements are more important when reconstructing an occurrence tensor. 
The error between the recovered low rank tensor $\tscp{\hat A,\hat B, \hat C}$ and the observation $\tsO$ 
grows with the noise ratio, while the error between $\tscp{\hat A, \hat B, \hat C}$ and the groundtruth $\tsX$ 
is relatively stable. 
This suggests that the recovered tensor $\tscp{\hat A, \hat B, \hat C}$ can be used to 
de-noise the observation. 

In \cref{fig:noise} (bottom), we set a $10\%$ noise ratio and vary the rank parameter $R$ for the recovered tensor $\tscp{\hat A, \hat B, \hat C}$. 
Note that $R$ is an upper bound of the CP rank of $\tscp{\hat A, \hat B, \hat C}$. 
NOTF could achieve zero for both false positives and false negatives when $R=6$, 
which means that the ground truth  can be completely recovered from the noisy observation. 
However, NOTF becomes unstable with larger $R$ and leads to large false negative counts. 
A possible reason is that the initialization by CP decomposition of $\tsO$ becomes less stable 
when large $R$ is used, and the least squares in \eqref{eq:sub2start}-\eqref{eq:sub2end} 
are often ill-posed and hard to solve.
We finally observe that modeling the sparse error by the $\ell_0$ norm brings an additional benefit
in that the recovered $\hat A, \hat B$, and $\hat C$ are sparse; this leads to a clearer interpretation 
of each rank-one tensor as a community.

\section{Experiments on the resMBS dataset}

We explore the roles played by financial institutions (FIs) across multiple contracts (FCs) 
using NOTF with the $\ell_0$ norm. ResMBS\cite{burdick2016dsmm,xu2016jdiq,xu2016dsmm} contains extracted relationship of FI (e.g., Bank of America) playing a role (e.g., issuer) for a specific financial contract. 
The discrete values of occurrence tensor $\tsO \in \bbR^{971 \times 85 \times 27}$ 
indicate the counts of extractions of the specific (FC, FI, Role)
occurrence from documents issued in 2005.  
$\tsO$ is sparse ($1.02 \%$ non-zero values) and extraction noise is estimated to be $ \leq 0.2\%$. 
We describe some observations here and present the relevant figures in \cref{app:exp} due to space limitations.

\begin{figure}[htbp]
\centerline{
\includegraphics[width= 0.9\linewidth]{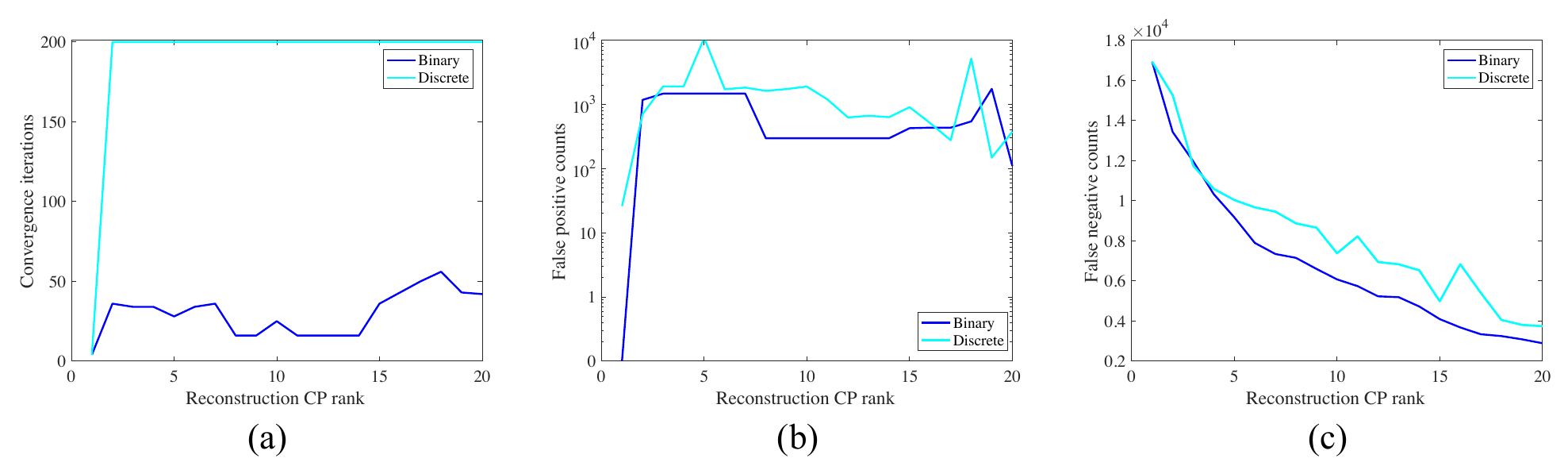}
}
\caption{\small  (a) Convergence iteration, (b) false positive counts, and (c) false negative counts when vary the tensor CP rank $R$ for the reconstruction of resMBS dataset. Both binary and discrete tensor of resMBS are tested. Note that reconstruction errors presented in (b)(c) are based on noisy observation $\tsO$ as the groundtruth is unknown. 
 }
\label{fig:resmbs}
\end{figure}

\begin{figure}[htbp]
\centerline{
\includegraphics[width= 0.4\linewidth]{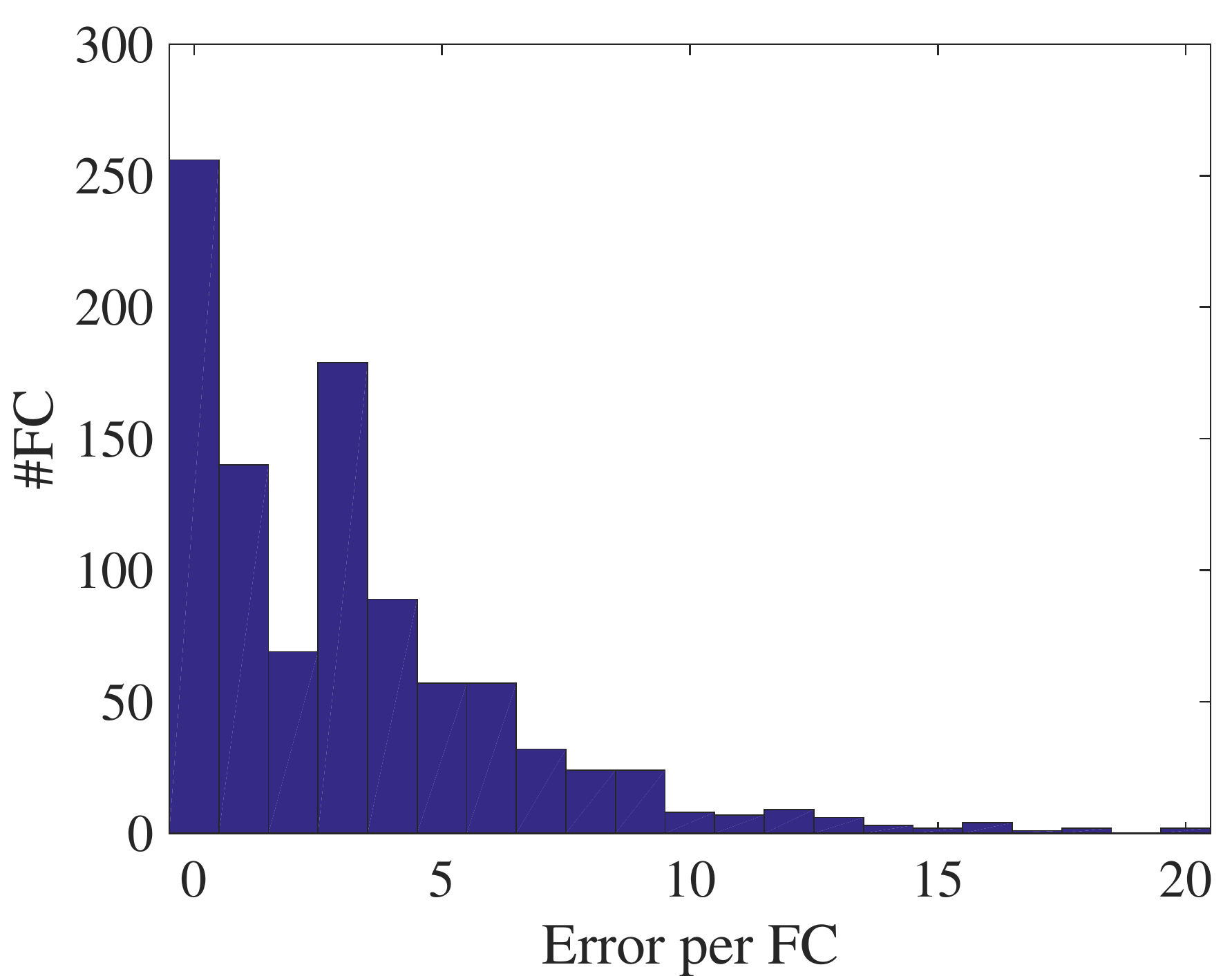}
\includegraphics[width= 0.4\linewidth]{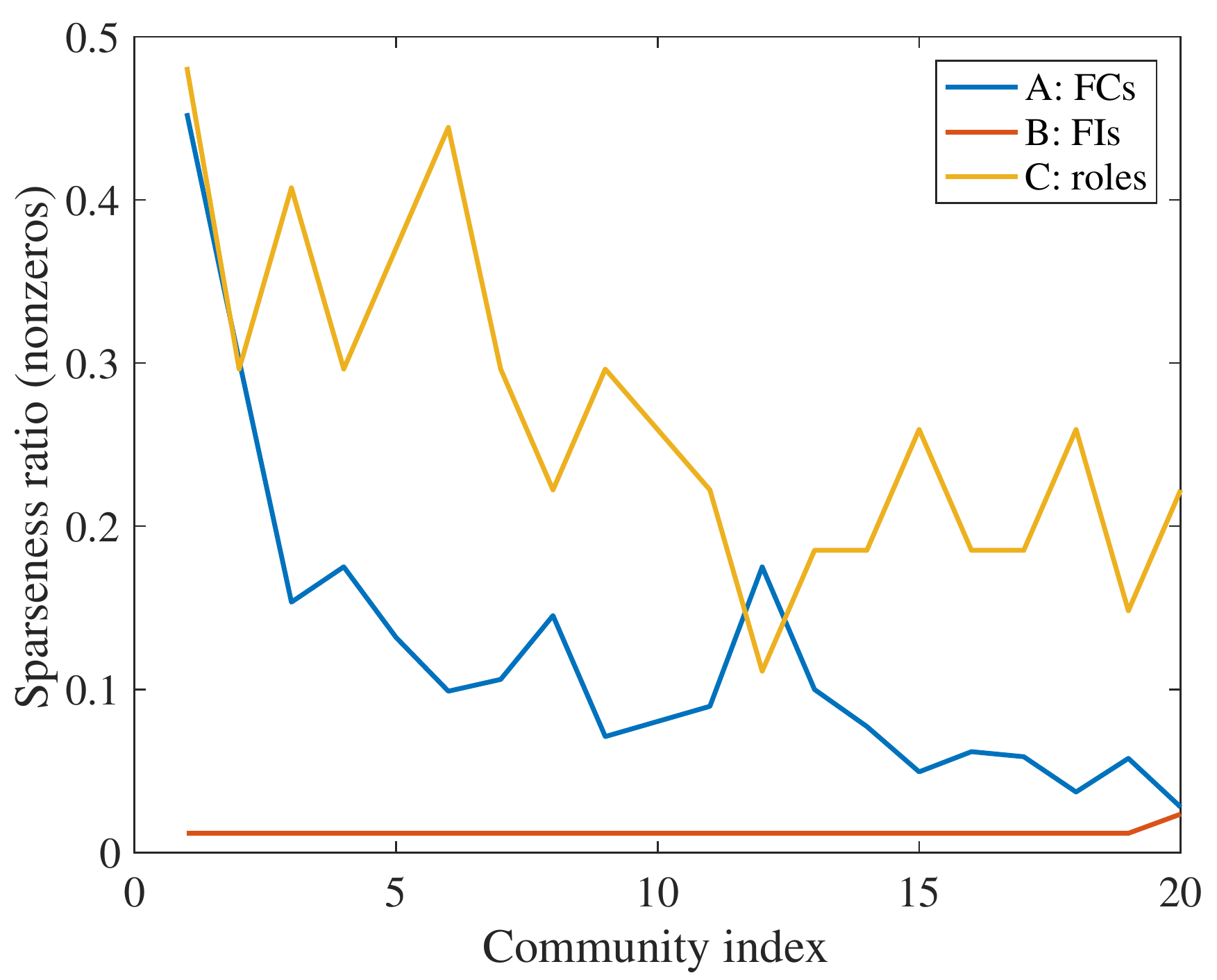}
}
\caption{\small (left) Histogram of errors when constructed resMBS tensor with $R=20$. (right) Nonzero ratio  for the constructed tensor component $\hat A, \hat B, \hat C$. Each rank-one tensor $a_r \circ b_r \circ c_r, r = 1,\ldots,R$ represents a community.
 }
\label{fig:hist}
\end{figure}

We vary the CP rank parameter $R$ and reconstruct tensors for both the discrete observation $\tsO$ 
and its binary version.
Performance is similar for both while it is notably slower to reconstruct discrete values (\cref{fig:resmbs}).We note that resMBS is challenging as the tensor is sparse. 
The false positives are relatively stable while the false negatives decrease as $R$ increases.  
With $R=20$, the total error count between the reconstructed tensor and the noisy observation is $3002$;
this roughly matches the expected errors of the information extractor. 
The histogram (\cref{fig:hist} (left)) shows that errors for each FC is in a reasonable range (0, 20)
with a mean of $3$.

\begin{figure}[thbp]
\centerline{
\includegraphics[width= \linewidth]{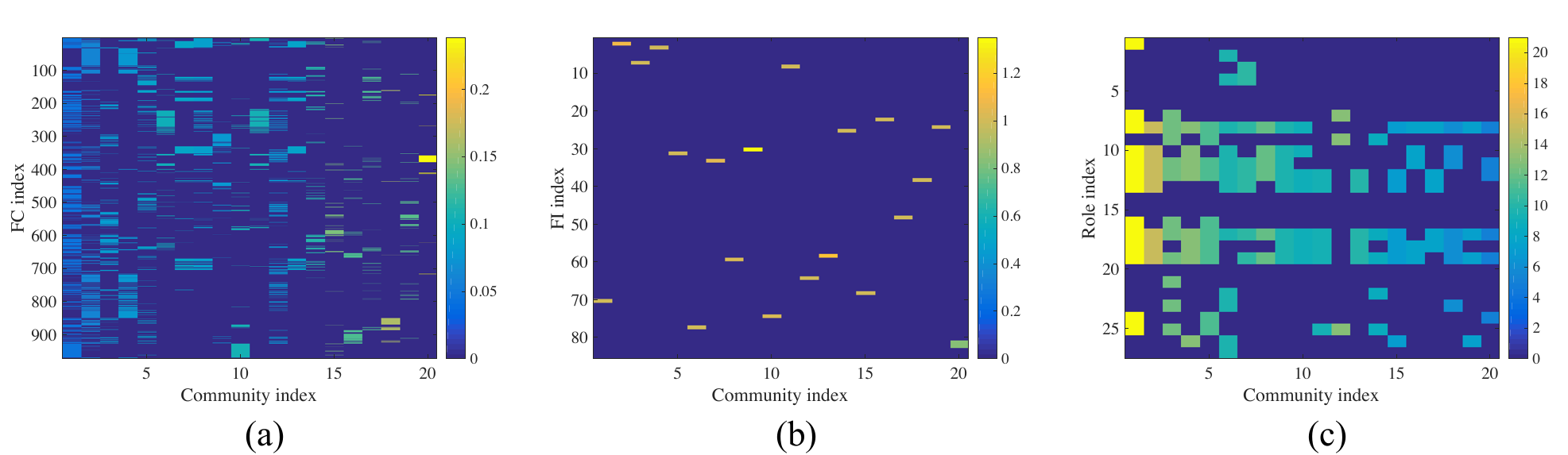}
}
\caption{\small The distribution of FC, FI, and roles in each community.
 }
\label{fig:sp2}
\end{figure}

At last, we examine the discovered communities by NOTF for resMBS.   Each rank-one tensor $a_r \circ b_r \circ c_r, r = 1,\ldots,R$ represents a community. \cref{fig:hist} (right) presents the nonzero ratio and \cref{fig:sp2} presents the distribution of the reconstructed tensor component $\hat A = [a_r], \hat B = [b_r], \hat C = [c_r]$. An interesting observation is that the communities are ``centered'' around FIs, i.e., each community only contains one or two FIs. Some FIs could play various roles and appear in various FCs, while some FIs only play a limited number of roles in a limited number of FCs.

\section{Discussion and future work}

We present non-negative occurrence tensor factorization (NOTF) for analyzing heterogeneous networks. 
CP tensor decomposition is adapted to discover the embedded communities. 
The $\ell_0$ norm is used to model the discrete tensor values and sparse errors, 
and the objective is solved with an efficient splitting optimization algorithm.
NOTF is applied to both synthetic data and a new heterogeneous 
bipartite graph, 
resMBS, 
representing financial role relationships extracted from financial contracts.
Preliminary results are promising and suggest that NOTF can be used to de-noise the occurrence tensors
and identify communities in resMBS. 

There are several directions for future work. 
The $\ell_0$ norm is known to be difficult to optimize. 
The $\ell_p$ norm ($0<p<1$) satisfies the KL inequality, is often used as a surrogate, and may provide a theoretical convergence guarantee. 
The penalty parameter $\tau$ is crucial for both convergence speed and solution quality for nonconvex problems;
adaptive ADMM\cite{xu2016adaptive,xu2016empirical} which automates the selection of $\tau,$ achieves promising practical performance. 
To deal with the high sparsity of the resMBS tensor, domain-specific constraints (e.g., 
{\it each FC should contain an FI play role ``Issuer''} \tom{I can't understand what this means because you never explain the dataset}) may boost performance. Finally, it is interesting to apply NOTF for analyzing some other heterogeneous networks that could be represented with an occurrence tensor. 


\subsubsection*{Acknowledgments}
ZX and TG were supported by US NSF grant CCF-1535902 and by US ONR grant N00014-15-1-2676.
ZX and LR were supported by NSF grants CNS1305368 and DBI1147144, and NIST award 70NANB15H194.

\small
\bibliographystyle{abbrvnat}
\bibliography{tensor,admm,resMBS}
\clearpage

\section{Appendix: experimental results for resMBS}
\label{app:exp}

\end{document}

%% file: definitions.tex
\usepackage{amsmath}
\usepackage{etoolbox}
\usepackage{bm}
\usepackage{mathtools}
\usepackage{algorithm}
\usepackage{algorithmic}
\usepackage{subfigure}
\usepackage{xspace}
\usepackage{multirow}
\usepackage[capitalise]{cleveref}
\usepackage{array}

\usepackage{footnote}
\makesavenoteenv{tabular}
\makesavenoteenv{table}

\crefname{myrule}{Rule}{Rule}

\newcommand{\MyMapTemplatePrefixc}[4]{\expandafter#1\csname#3#4\endcsname{#2{#4}}} 
\forcsvlist{\MyMapTemplatePrefixc {\def} {\mathcal}{mc}} {A,B,C,D,E,F,G,H,I,J,K,L,M,N,O,P,Q,R,S,T,U,V,W,X,Y,Z}  
\usepackage[mathscr]{eucal}
\forcsvlist{\MyMapTemplatePrefixc {\def}{\mathscr}{ts}} {A,B,C,D,E,F,G,H,I,J,K,L,M,N,O,P,Q,R,S,T,U,V,W,X,Y,Z} 

\newcommand{\MyMapTemplatePrefixtb}[5]{\expandafter#1\csname#4#5\endcsname{#2{#3{#5}}}} 
\forcsvlist{\MyMapTemplatePrefixtb {\def} {\tilde}{\mathbf}{t}} {A,B,C,D,E,F,G,H,I,J,K,L,M,N,O,P,Q,R,S,T,U,V,W,X,Y,Z}  
\forcsvlist{\MyMapTemplatePrefixtb {\def} {\tilde}{\mathbf}{t}} {0,1,a,b,c,d,e,f,g,h,i,j,k,l,m,n,o,p,q,r,s,t,u,v,w,x,y,z}  

\newcommand{\MyMapTemplateNoPrefix}[3]{\expandafter#1\csname#3\endcsname{#2{#3}}}
\forcsvlist{\MyMapTemplateNoPrefix {\def} {\mathbf} } {0,1,a,b,c,d,e, f, g, h, i, j, k, l, m, n, o, p, q, r, s, t, u, v, w, x, y, z} 
\forcsvlist{\MyMapTemplateNoPrefix {\def} {\mathbf} } {A,B,C,D,E,F,G,H,I,J,K,L,M,N,O,P,Q,R,S,T,U,V,W,X,Y,Z}  

\def\gge{\epsilon}

\def\gl{\lambda}

\def\bbR{{\mathbb R}}

\def\st{\mbox{subject to}}

\newcommand{\trans}{^T}

\newcommand{\tscp}[1]{[\![ #1 ]\!]}

\newcommand{\zheng}[1]{}

%% file: nips_2016_v3.bbl
\begin{thebibliography}{22}
\providecommand{\natexlab}[1]{#1}
\providecommand{\url}[1]{\texttt{#1}}
\expandafter\ifx\csname urlstyle\endcsname\relax
  \providecommand{\doi}[1]{doi: #1}\else
  \providecommand{\doi}{doi: \begingroup \urlstyle{rm}\Url}\fi

\bibitem[Anandkumar et~al.(2014)Anandkumar, Ge, Hsu, Kakade, and
  Telgarsky]{anandkumar2014tensor}
A.~Anandkumar, R.~Ge, D.~Hsu, S.~M. Kakade, and M.~Telgarsky.
\newblock Tensor decompositions for learning latent variable models.
\newblock \emph{Journal of Machine Learning Research}, 15\penalty0
  (1):\penalty0 2773--2832, 2014.

\bibitem[Anandkumar et~al.(2015)Anandkumar, Jain, Shi, and
  Niranjan]{anandkumar2015tensor}
A.~Anandkumar, P.~Jain, Y.~Shi, and U.~Niranjan.
\newblock Tensor vs matrix methods: Robust tensor decomposition under block
  sparse perturbations.
\newblock \emph{arXiv preprint}, 2015.

\bibitem[Bader and Kolda(2006)]{TTB_Dense}
B.~W. Bader and T.~G. Kolda.
\newblock Algorithm 862: {MATLAB} tensor classes for fast algorithm
  prototyping.
\newblock \emph{ACM Transactions on Mathematical Software}, 32\penalty0
  (4):\penalty0 635--653, December 2006.
\newblock \doi{10.1145/1186785.1186794}.

\bibitem[Boyd et~al.(2011)Boyd, Parikh, Chu, Peleato, and
  Eckstein]{boyd2011admm}
S.~Boyd, N.~Parikh, E.~Chu, B.~Peleato, and J.~Eckstein.
\newblock Distributed optimization and statistical learning via the alternating
  direction method of multipliers.
\newblock \emph{Found. and Trends in Mach. Learning}, 3:\penalty0 1--122, 2011.

\bibitem[Burdick et~al.(2016)Burdick, De, Raschid, Shao, Xu, and
  Zotkina]{burdick2016dsmm}
D.~Burdick, S.~De, L.~Raschid, M.~Shao, Z.~Xu, and E.~Zotkina.
\newblock {resMBS}: Constructing a financial supply chain graph from financial
  prospecti.
\newblock In \emph{SIGMOD DSMM workshop}. ACM, 2016.

\bibitem[Carroll and Chang(1970)]{carroll1970analysis}
J.~D. Carroll and J.-J. Chang.
\newblock Analysis of individual differences in multidimensional scaling via an
  n-way generalization of “eckart-young” decomposition.
\newblock \emph{Psychometrika}, 35\penalty0 (3):\penalty0 283--319, 1970.

\bibitem[Chen et~al.(2016)Chen, Han, Wang, Zhao, Meng, and
  Tang]{chen2016robust}
X.~Chen, Z.~Han, Y.~Wang, Q.~Zhao, D.~Meng, and Y.~Tang.
\newblock Robust tensor factorization with unknown noise.
\newblock In \emph{Proceedings of the IEEE Conference on Computer Vision and
  Pattern Recognition}, pages 5213--5221, 2016.

\bibitem[Goldfarb and Qin(2014)]{goldfarb2014robust}
D.~Goldfarb and Z.~Qin.
\newblock Robust low-rank tensor recovery: Models and algorithms.
\newblock \emph{SIAM Journal on Matrix Analysis and Applications}, 35\penalty0
  (1):\penalty0 225--253, 2014.

\bibitem[Goldstein and Osher(2009)]{goldstein2009split}
T.~Goldstein and S.~Osher.
\newblock The split {B}regman method for {L}1-regularized problems.
\newblock \emph{SIAM Journal on Imaging Sciences}, 2\penalty0 (2):\penalty0
  323--343, 2009.

\bibitem[Goldstein et~al.(2014)Goldstein, O'Donoghue, Setzer, and
  Baraniuk]{goldstein2014fast}
T.~Goldstein, B.~O'Donoghue, S.~Setzer, and R.~Baraniuk.
\newblock Fast alternating direction optimization methods.
\newblock \emph{SIAM Journal on Imaging Sciences}, 7\penalty0 (3):\penalty0
  1588--1623, 2014.

\bibitem[Gu et~al.(2014)Gu, Gui, and Han]{gu2014robust}
Q.~Gu, H.~Gui, and J.~Han.
\newblock Robust tensor decomposition with gross corruption.
\newblock In \emph{Advances in Neural Information Processing Systems}, pages
  1422--1430, 2014.

\bibitem[Harshman(1970)]{harshman1970foundations}
R.~A. Harshman.
\newblock Foundations of the parafac procedure: Models and conditions for an"
  explanatory" multi-modal factor analysis.
\newblock 1970.

\bibitem[Huang and Ding(2008)]{huang2008robust}
H.~Huang and C.~Ding.
\newblock Robust tensor factorization using r 1 norm.
\newblock In \emph{Computer Vision and Pattern Recognition, 2008. CVPR 2008.
  IEEE Conference on}, pages 1--8. IEEE, 2008.

\bibitem[Kolda and Bader(2009)]{kolda2009tensor}
T.~G. Kolda and B.~W. Bader.
\newblock Tensor decompositions and applications.
\newblock \emph{SIAM review}, 51\penalty0 (3):\penalty0 455--500, 2009.

\bibitem[Liavas and Sidiropoulos(2015)]{liavas2015parallel}
A.~P. Liavas and N.~D. Sidiropoulos.
\newblock Parallel algorithms for constrained tensor factorization via
  alternating direction method of multipliers.
\newblock \emph{IEEE Transactions on Signal Processing}, 63\penalty0
  (20):\penalty0 5450--5463, 2015.

\bibitem[Maruhashi et~al.(2011)Maruhashi, Guo, and
  Faloutsos]{maruhashi2011multiaspectforensics}
K.~Maruhashi, F.~Guo, and C.~Faloutsos.
\newblock Multiaspectforensics: Pattern mining on large-scale heterogeneous
  networks with tensor analysis.
\newblock In \emph{Advances in Social Networks Analysis and Mining (ASONAM),
  2011 International Conference on}, pages 203--210. IEEE, 2011.

\bibitem[Papalexakis et~al.(2016)Papalexakis, Faloutsos, and
  Sidiropoulos]{papalexakis2016tensors}
E.~E. Papalexakis, C.~Faloutsos, and N.~D. Sidiropoulos.
\newblock Tensors for data mining and data fusion: Models, applications, and
  scalable algorithms.
\newblock \emph{ACM Transactions on Intelligent Systems and Technology (TIST)},
  8\penalty0 (2):\penalty0 16, 2016.

\bibitem[Schein et~al.(2015)Schein, Paisley, Blei, and
  Wallach]{schein2015bayesian}
A.~Schein, J.~Paisley, D.~M. Blei, and H.~Wallach.
\newblock Bayesian poisson tensor factorization for inferring multilateral
  relations from sparse dyadic event counts.
\newblock In \emph{Proceedings of the 21th ACM SIGKDD International Conference
  on Knowledge Discovery and Data Mining}, pages 1045--1054. ACM, 2015.

\bibitem[Xu and Raschid(2016)]{xu2016dsmm}
Z.~Xu and L.~Raschid.
\newblock Probabilistic financial community models with latent dirichlet
  allocation for financial supply chains.
\newblock In \emph{SIGMOD DSMM workshop}. ACM, 2016.

\bibitem[Xu et~al.(2016{\natexlab{a}})Xu, Burdick, and Raschid]{xu2016jdiq}
Z.~Xu, D.~Burdick, and L.~Raschid.
\newblock Exploiting lists of names for named entity identification of
  financial institutions from unstructured documents.
\newblock \emph{arXiv preprint arXiv:1602.04427}, 2016{\natexlab{a}}.

\bibitem[Xu et~al.(2016{\natexlab{b}})Xu, De, Figueiredo, Studer, and
  Goldstein]{xu2016empirical}
Z.~Xu, S.~De, M.~A.~T. Figueiredo, C.~Studer, and T.~Goldstein.
\newblock An empirical study of admm for nonconvex problems.
\newblock In \emph{NIPS workshop on nonconvex optimization},
  2016{\natexlab{b}}.

\bibitem[Xu et~al.(2016{\natexlab{c}})Xu, Figueiredo, and
  Goldstein]{xu2016adaptive}
Z.~Xu, M.~A. Figueiredo, and T.~Goldstein.
\newblock Adaptive {ADMM} with spectral penalty parameter selection.
\newblock \emph{arXiv preprint arXiv:1605.07246}, 2016{\natexlab{c}}.

\end{thebibliography}
